\documentclass[aps,onecolumn, 
pre,superscriptaddress, longbibliography, 
notitlepage]{revtex4-2}
\usepackage{bm,amsmath,graphicx, amssymb, mathtools,multirow}
\usepackage[colorlinks=true,linkcolor=MidnightBlue,urlcolor=black,citecolor=MidnightBlue,anchorcolor=MidnightBlue]{hyperref}
\usepackage[dvipsnames]{xcolor}
\begin{document}
%%%%%====================================
%%%%%====================================

\title{Flocking transition in phoretically interacting active particles with pinning disorder}
\author{Sagarika Adhikary}
\email{a.sagarika@physics.iitm.ac.in}
\affiliation{Department of Physics, Indian Institute of Technology Madras, Chennai 600036, India}
\author{Arvin Gopal Subramaniam}
\affiliation{Department of Physics, Indian Institute of Technology Madras, Chennai 600036, India}
\author{Rajesh Singh}
\affiliation{Department of Physics, Indian Institute of Technology Madras, Chennai 600036, India}

\begin{abstract}
Recent studies in the collective behavior of active colloids have shown that a global polar order may emerge due to long-ranged chemo-repulsive interactions between them. Here, we report the role of pinning disorder in the flocking transition for such a system. To this end, we study the problem of chemically interacting active colloids with some fraction of the colloids randomly pinned over space such that they can only rotate while phoretically interacting with other particles. Using this model, we investigate the sustenance of global polar order in the presence of quenched spatial disorder. We quantify the flocking transition by studying the global polarization, and the role of finite-size effects.
We find that in the crystallite flocking phase, even a small fraction of pinning can destroy spatial crystalline order, although polar order in the form of a liquid phase is maintained. It is observed that polar order is sustained in a system with a higher pinning fraction if the long-ranged repulsive force is subsequently increased. However, in absence of chemo-repulsive forces between particles, polar order drastically decreases even with a smaller pinning fraction. Our work suggests that the flocking transition of active colloids can be controlled via "translationally inert" obstacles, that rotate but do not translate whilst interacting with the bulk. 
\end{abstract}

\maketitle
\maketitle
\section{Introduction}\label{sec:intro}
Active matter is a class of non-equilibrium systems composed of many individual components that each consume energy from their environment and convert it into self-propulsion or mechanical work \cite{vrugt2025exactly}. The individual components of active matter are referred to as active particles, whose examples include autophoretic colloids \cite{ebbens2010pursuit} and microorganisms \cite{goldstein2015green}. 
Many-body interactions between active particles give rise to emergent phenomena such as phase separation \cite{Cates_2025, cates2015motility, marchetti2016minimal}, 
flocking \cite{toner2024physics, toner2005hydrodynamics, marchetti2013hydrodynamics} and other collective motion \cite{bowick2022symmetry, vicsek2012collective, wensink2012emergent, schaller2010polar, miranda2025collective, knevzevic2022collective, giavazzi2018flocking, chandrasekaran2019percolation, grossmann2020particle}. Numerous studies have investigated the collective dynamics of active matter systems, focusing on various types of interactions such as velocity-alignment mechanisms \cite{chen2024emergent, martin2018collective, sese2018velocity}, non-reciprocity \cite{saha2025effervescence,kreienkamp2024dynamical}, history-dependent interactions \cite{kumar2023emergent, subramaniam2024rigid, horton2025order}, and time-delayed interactions \cite{horton2025order, piwowarczyk2019influence, erban2016cucker}. The study of how an emergent polar order can arise from interacting active particles without alignment interactions \cite{vicsek1995novel} are numerous, and consist of several inertial models \cite{caprini2023flocking, grossman2008emergence, hanke2013understanding, miranda2025collective}, short-ranged interactions (cognitive/biophysical based models) \cite{hiraiwa2019two, hiraiwa2020dynamic, grossmann2013self}, and long-ranged repulsive interactions \cite{knevzevic2022collective, subramaniam2025minimal, das2024flocking}. \\

Despite these advances, a comprehensive understanding of how different types of interactions and disorders give rise to collective motion in complex environments remains to be explored more extensively \cite{bechinger2016active,gompper20252025,rahmani2020flocking}. 
Recently, in a model with long-range chemo-repulsive interactions, a flocking transition was observed even without explicit alignment interactions among the particles, where a flocking transition was shown to occur exclusively via the incorporation of short-ranged excluded volume repulsion and long-ranged repulsive torques \cite{subramaniam2025minimal}. Most studies of self-propelled particles, however, are typically conducted in homogeneous environments with monodisperse particles. In contrast, real active systems often exhibit complexity and heterogeneity. Various studies incorporating complexities such as presence of various disorder \cite{vahabli2023emergence,morin2017distortion,stoop2018clogging,peruani2018cold, sandor2017dynamic,olsen2021active}, quenched disorder \cite{duan2021breakdown,das2018polar}, obstacles or perturbation \cite{codina2022small, mokhtari2017collective, adhikary2021effect}, geometric constraints \cite{murali2022geometric}, population heterogeneity \cite{adhikary2022pattern, rouzaire2025activity,negi2025binary,tang2025reentrant} have been explored. 
Understanding how collective motion emerges and persists under complexities such as geometrical disorder remains a fundamental and intriguing problem. 

In this work, we study a system of active particles with chemo-repulsive forces and torques in the presence of quenched disorder. In our model, some fraction of the active particles are randomly pinned over the two-dimensional space. The pinned particles can only rotate and do not translate. However, the interactions between the pinned and free particles remain the same. We focus on understanding the role of pinning sites in affecting the flocking transitions.
We find that increasing the density of pinning sites leads to a systematic reduction in the global polar order. In the absence of disorder, the system exhibits a crystallite flocking phase; however, even a small degree of disorder disrupts this crystallite structure, giving rise instead to a liquid-like flocking state. As the pinning strength or density is further increased, the system transitions into a randomly disordered phase. In a case without long-ranged repulsive force, up to a relatively smaller fraction of pinning, polar order sustains in the system. We construct comprehensive phase diagrams mapping the behavior of the system in terms of global polar order, susceptibility, density variance, and hexatic order parameter. Additionally, we analyze the flocking transition, pair correlation functions, and other steady-state collective properties to elucidate the interplay between disorder, interactions, and emergent order in active systems.\\

\begin{figure*}[t]
    \centering
    \includegraphics[width=0.8\textwidth]{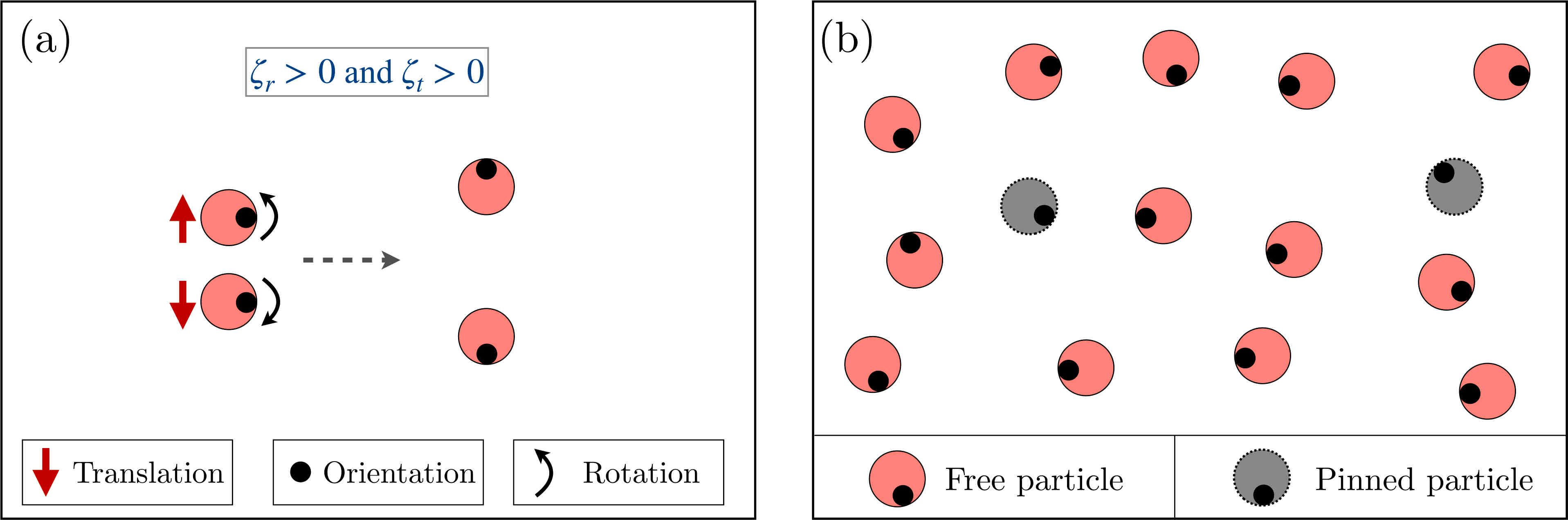}
    \caption{\label{model}
        (a) Schematic for the interactions and dynamics of two active colloids (shown as orange circles with a black dot to indicate orientation).  Solid red arrows show chemo-repulsive forces between particles while black curved arrows denote chemo-repulsive torques. Schematic of the model of active colloids with random pinning is show in (b). Pinned particles (grey colour) only rotate (no translational motion for pinned particles), while active particles rotate as well as translate.}
\label{fig:model}
\end{figure*}

The remainder of the paper is organized as follows. In section \ref{sec:Model}, we describe our model of phoretically interacting active particles in the presence of pinning disorder. In section  \ref{sec:results}, we present our results from particle-based numerical simulations by mapping phase diagrams and describing the properties of the different phases. First, this is presented for the general case both long-ranged forces and torques present in the system, where various collective properties of the system and the robustness of the polar order to pinning discussed. Then in section \ref{sec:v}, we study the case without long-ranged repulsive forces. 
%We also studied a case without short-ranged steric repulsion in appendix \ref{sec:ap1}. 
Finally in section \ref{sec:concEv}, we summarize the main findings, compare them with existing literature, and outline potential directions for future research. \\

\section{Model}\label{sec:Model}
%%%=============================================
%%%=============================================
\subsection{Equations of motion for active particles}
We consider a set of $N$ chemically interacting active colloids of radius $b$ in two-dimensions. 
Of these $N$ particles, $N_m$ are mobile and $N_p$ are pinned, such that $N_m+N_p=N$.
A schematic describing their motion is shown in Fig. \ref{fig:model}.
The $N_p$ randomly pinned particles can only rotate, and does not have any translational motion. Simultaneously, the other $N_m$ moving particles can both translate and rotate. 
We model the $i$th active particle as a colloid particle centered at $\mathbf r_i=(x_i,y_i)$, confined to move in two-dimensions, which self-propels with a speed $v_s$,  along the directions $\mathbf e_i =(\cos\theta_i,\,\sin\theta_i)$. Here $i=1,2,3,\dots,N$ and $\theta_i$ is the angle made by the orientation $\mathbf e_i$ with respect to the positive $x$-axis.
The dynamics of $i$th particle follows:
\begin{align}
\label{eq:mainLE}
     \dot{\mathbf r}_i  &=  v_s \mathbf e_i +\zeta_t\, \mathbf J_i + \mu \mathbf F_i,
     \qquad \qquad \,\,i=1,\dots N_m,\\ 
     \dot{ \mathbf e}_i&= \left[\zeta_r 
    \left(\mathbf e_i\times\mathbf {\mathbf{{J}}}_i 
    \right) + \bm \eta^r_i\right] \times \mathbf e_i    ,\,\, \qquad \forall i.\nonumber
\end{align}
Here $v_s$ is the self-propulsion speed of an isolated moving particle. On the other hand, the pinned particles can not move such that:    
$\dot {\mathbf r}_{i} = 0$
for
$i=N_m+1,\dots , N_m+N_p$.
In the above, $\bm \eta^r_i$ is a white noise with zero mean and $2D_r$ variance with no temporal correlation. 
As shown below in section \ref{sec:concEv}, the phoretic flux $\mathbf{J}_i(t)$ (of chemical origin) is obtained in terms of a scalar field $c(\mathbf r, t)$, 
which is the concentration of the chemicals (e.g filled micelles in an oil-emulsion system \cite{hokmabad2022chemotactic, kumar2023emergent}). 
A short-ranged repulsive interaction is added in Eq.\eqref{eq:mainLE} to preclude overlap of the particles through the body force: 
$\mathbf F_i = - \boldsymbol{\nabla}_i \mathcal{U}$. Here, $\mathcal{U}= \sum_{i<j}\mathcal U^e   (\mathbf r_i,\mathbf r_{j})$. With $b$ as the colloidal radius, we choose $\mathcal U^e$ to be of the form:
$\mathcal U^e=\kappa\left(r_{ij}-2b\right)^2$ if $|\mathbf{r}_{i} - \mathbf{r}_{j}|<2b$, while it vanishes otherwise. 
Here,
$\kappa$ is a constant which determines the strength of the repulsive force.

\subsection{Dynamics of the phoretic field}\label{sec:concEv}
In Eq.\eqref{eq:mainLE}, the flux $\mathbf{J}_i(t) 
$ is given in terms of the phoretic (chemical) field $c(\mathbf r, t)$.
The phoretic field is obtained by considering each particle as an emitter of the chemical field with emission constant $\lambda_0$, while the field decays uniformly at a rate $\lambda_d$.
Consequently, the phoretic field $c(\mathbf r, t)$ evolves in time as
\begin{align}
     D_{c}\nabla^2 c(\mathbf r, t) + \sum_{i=1}^N\lambda_0\, \delta(\mathbf r- \mathbf r_i)
-\lambda_d c= 0.
\end{align}
 Here, $D_{c}$ is the diffusion coefficient. We note that and we are working in the limit of instantaneous chemical interactions between the colloidal surfaces \cite{saha2014clusters, soto2015self,ruckner2007chemically,singh2019competing}. Thus, we work in the limit that of steady-state diffusion equation for the chemical field. 
The expression of the phoretic flux $\mathbf J$ and phoretic field $c(\mathbf r)$ then follows: 
\begin{align}
 \mathbf{J}_i(t) =-
 \left[\mathbf\nabla c(\mathbf r, t)\right]_{\mathbf r=\mathbf r_i} = \frac{\lambda_0}{4\pi D_c }\sum^N_{\substack{j=1\\ i\neq j}} 
 \frac{\mathbf  r_{ij} \exp (-\lambda r_{ij})(\lambda r_{ij}+1)}{
 r_{ij}^3 },
 \qquad    c(\mathbf{r}, t) = \frac{\lambda_0}{4\pi D_c }\sum^N_{{j=1
    }} 
 \frac{ \exp\left[-\lambda\left(\mathbf{r} - \mathbf{r}_{j}\right) \right]
}
 {|\mathbf{r} - \mathbf{r}_{j}|} 
 .
 \label{eq:curr_r2}
\end{align}
Here, $\lambda=\sqrt{\lambda_d/D_c}$. 
 Using the obtained expression of flux, 
it follows that the term
$\zeta_r$ in Eq.(\ref{eq:mainLE}) turns particles away from each other if $\zeta_r>0$ through interparticle chemical interactions. Similarly, the term proportional to $\zeta_t$ ensures repulsion in the positional dynamics if $\zeta_t>0$. In this paper, both $\zeta_r$ and $\zeta_t$ are positive and the system is said to be \textit{chemo-repulsive}. \\

\subsection{Dimensionless parameters}
We first enumerate some dimensionless quantities to study the system. 
An important parameter in this model is the pinning fraction $n_p$, which is defined as 
% We define the pinning fraction $n_p$ as:
\begin{align}
    n_p = \frac{N_p}{N_m+N_p} =  \frac{N_p}{N} .
    %=  \frac{1}{N_m/N_p+1}  
    \label{eq:pin}
\end{align}
Further, we define two important dimensionless parameters:
\begin{align}
    \Lambda_{r} &=\frac{\tau}{\tau_r}= \frac{\zeta_r }{b^3v_s}, \qquad\quad
    \Lambda_{t} = \frac{\zeta_t}{b^4 v_s}.
\label{eq:lscales}
\end{align} 
Here, $\tau = {b}/{v_s}$ is the propulsion time scale of the moving particles, while
$\tau_r=b^4/\zeta_r$, characterizes the time scale associated with deterministic rotational motion of the particle arising from phoretic interactions.  
We note that in our model, the chemical field diffuses in an infinite three-dimensional half-space bounded by a plane surface. 
But, the particles only move in two-dimensions, such that they move in the plane of the bounding surface. 
Thus, 
the dimensions of chemical concentration is: 
$[c]=[\text{L}^{-3} ]$, while the dimensions of the
chemical flux are $[\mathbf J]=[\text{L}^{-4} ]$.
So, the dimensions of parameters are:
$[\zeta_r]=[\text{L}^{4}\text{T}^{-1} ]$ 
and $[\zeta_t]=[\text{L}^{5}\text{T}^{-1} ]$.\\

In the following, we use these three parameters $n_p$, $\Lambda_r$ and $\Lambda_t$  to study the global polar order, fluctuations in order parameter, density variance, hexatic order and other collective properties of the model.\\

\subsection{Simulation details}
We simulate Eq. \eqref{eq:mainLE} using an Euler-Maruyama integrator. Initially, all the moving and pinned particles are randomly distributed over the two-dimensional space. The initial orientations are also randomly chosen from the uniform distribution of angles [$-\pi,+\pi$] (the angle is determined with respect to the positive $x$-axis). Then, pinned particles are randomly chosen and the corresponding collective dynamics is studied. Periodic boundary conditions are applied on both the $x$-axis and $y$-axis of a square box of length $L$. The area fraction $\phi$ is related to the number density $\rho$, as $\phi= {\left(N\pi b^2\right)}/{L^2}=\rho \pi b^2$. The area fraction is fixed at $\phi=0.36$ for all the simulations, except only in the phase transition and finite-size effect part. See appendix \ref{app:sim} for a table of parameters. \\

%%%=============================================
%%%=============================================
\section{Results}\label{sec:results}
%%%=============================================
%%%=============================================
\begin{figure*}[t]
    \centering
    \includegraphics[width=\textwidth]{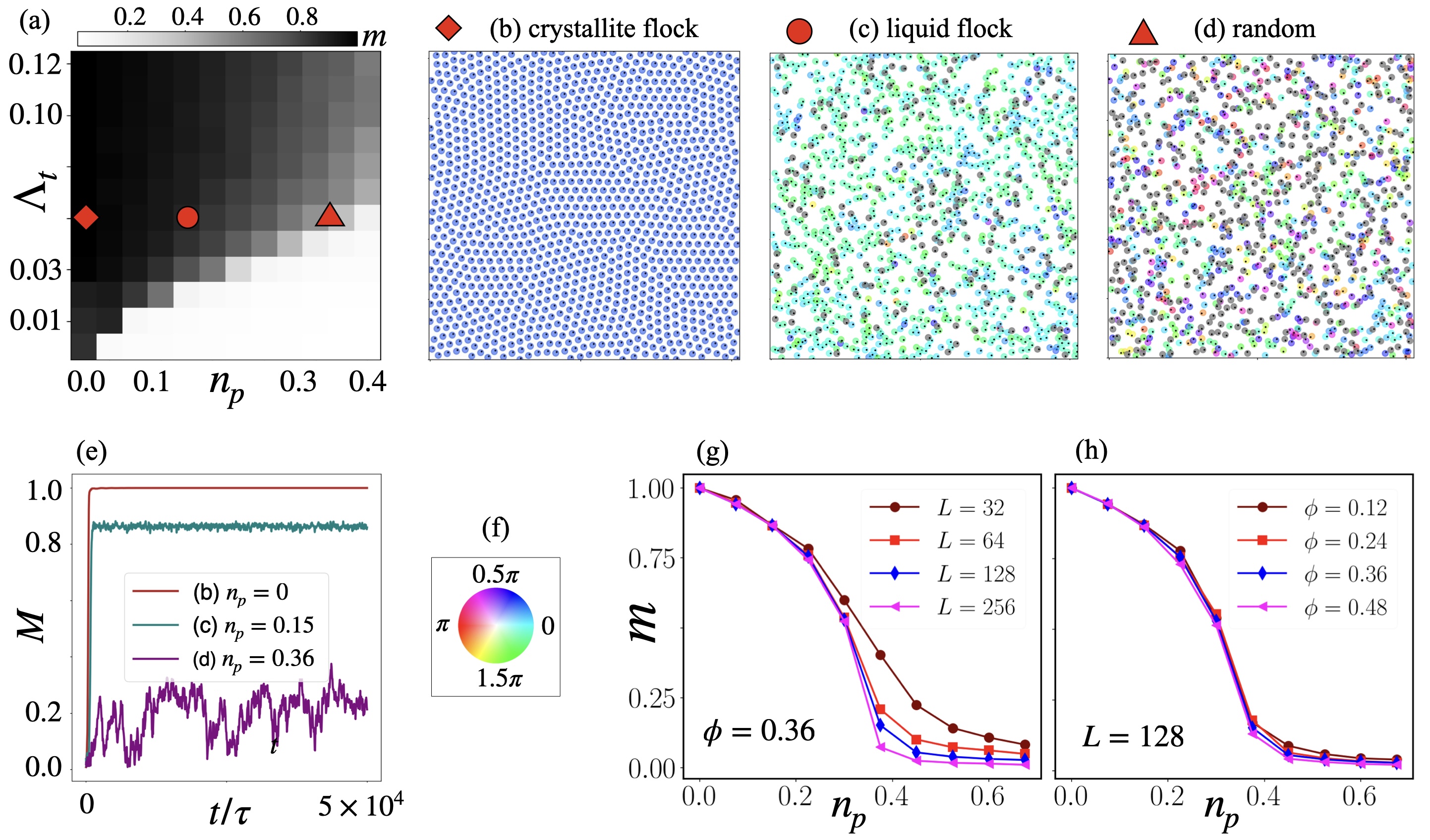}
    \caption{\label{phase_D}
    (a) Phase diagram in the $(n_p,\Lambda_t)$ plane, constructed via the order parameter $m$; three reference points on the diagram correspond to the phases are shown in (b) crystallite flock ($n_p=0$), (c) polar liquid ($n_p=0.15$) and (d) random ($n_p=0.36$). The total number of particles are taken as $N=1200$. Pinned particles are marked in grey colour and direction of motion of the free particles follows the color scheme shown in (f); (e) Polarization $M$ versus time for (b)-(d) phases; (f) color scheme of moving particles, the angle is determined with respect to positive $x$-axis.  
    (g) Order parameter $m$ versus $n_p$ for different system of sizes $L=32,64,128$ and $256$ (constant $\phi=0.36$). (h) Order parameter $m$ versus $n_p$ for different area fraction $\phi=0.12,0.24,0.36$ and $0.48$ ($L=128$ is constant).
    }
\label{fig:phase_D1}
\end{figure*}
%%%=============================================
\subsection{Effect of pinning disorder on flocking}
The global polarization \cite{vicsek1995novel,subramaniam2025minimal,adhikary2022pattern} of the system can be defined by the order parameter $m$ as:
\begin{align}
    m = \langle M \rangle_{ss},\qquad     M = \bigg| \frac1N \sum_{i=1}^N \mathbf{e}_i \bigg|
\end{align}

\begin{figure*}[t]
    \centering
    \includegraphics[width=0.98\textwidth]{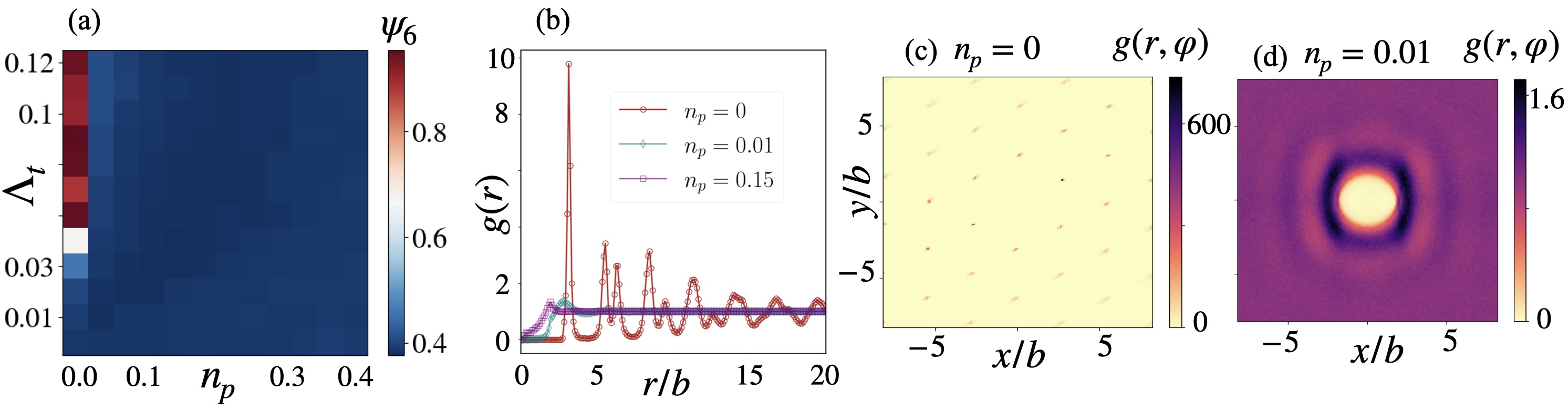}
    \caption{ 
    (a) Phase diagram of hexatic order in the $(n_p,\Lambda_t)$ plane. (b) Pair correlation $g(r)$ versus distance ($r$) plot for different pinning fraction $n_p=0$, $0.01$ and $0.15$. 
(c) $g(r, \varphi)$ plot with $n_p=0$, indicating a crystallite flock structure, (d) $g(r, \varphi)$ plot with $n_p=0.01$ indicating a liquid flock structure. %$N=4000$.
Here, the total number of particles, $N=4000$.}
\label{fig:pis6GR}
\end{figure*}

Here, the angular brackets denote averaging over the steady state. For this, the first $10^5$ time steps are excluded to reach the steady-state and then the next $10^5$ time steps are taken for time averaging (for each $8$ different realizations or initial configurations). Thus, a total size of $8 \times 10^5$ configurations 
are taken for statistical averages throughout this paper, unless specified otherwise. The instantaneous mean polarization $M$ at each time step is given by the average orientation of all particles.
The global polarization $m$ serves as an order parameter characterizing the flocking transition in terms of orientational order.
A phase diagram in the ($n_p$,$\Lambda_t$) plane is presented in Fig.\ref{fig:phase_D1}(a), showing the variation of the polar order $m$ across different parameter regimes. For this analysis, the repulsive torque strength is fixed at $\Lambda_r=0.1$. As we increase the pinning fraction, a higher value of $\Lambda_t$ (translational repulsion) is needed to maintain the ordered phase. Three representative points corresponding to distinct phases are indicated in Fig.\ref{fig:phase_D1}(a) and illustrated in the accompanying snapshots in Fig.\ref{fig:phase_D1}(b), (c) and (d) respectively. The particle orientations are color-coded according to their angle with respect to the positive $x$-direction, as shown in Fig.\ref{fig:phase_D1}(f). The three observed phases are as follows: (b) the first one is a  crystallite flock (where the term "crystallite" implies that this state has a finite-size crystalline order) without any pinning ($n_p=0$), exhibiting both orientational and structural order, as confirmed by the high global hexatic order (also refer to Fig.\ref{fig:pis6GR}(a)); (c) the second one is a polar liquid phase with pinning fraction $n_p=0.15$, where the hexatic order is lost and only polar order persists; (d) random or disordered phase with higher pinning fraction $n_p=0.36$, located near the flocking transition, where particle motion becomes random.
The temporal evolution of the polarization for these phases is shown in Fig.\ref{fig:phase_D1}(e). Starting from a random initial configuration, the polarization starts from a zero value and reaches its steady-state after some transient time. For a fixed $\Lambda_t$, increasing the pinning fraction results in a decrease in the steady-state polarization and an increase in fluctuations of $M$. For $n_p=0.36$, near the transition, the fluctuations in the order parameter are significantly larger than the other phases. These fluctuations are analyzed in detail later in the text to characterize additional collective properties of the system.\\

%%%=============================================
%%%=============================================
\subsection{Phase transition and Finite size effects}
Now we study phase transition and finite-size scaling for the order-disorder flocking transition for fixed values of $\Lambda_r=0.1, \Lambda_t=0.05$. As the system evolves from a random initial configuration to a steady state, the polar order parameter $m(n_p,L)$ is evaluated by taking its time average, where $L$ denotes the system size. In the ordered phase (small $n_p$), when all particles move coherently in the same direction, the order parameter $m$ approaches to $1$. Conversely, in the disordered phase (large $n_p$), where particle motion is random, the order parameter $m$ tends towards $0$. There exists a critical value of $n_p$, at which the system undergoes a transition between these two phases. The phase transition plot is shown in Fig.\ref{fig:phase_D1}(g) for different system of sizes, ranges from $L=32$ to $256$ (with fixed area fraction $\phi=0.36$). As the system size increases, finite-size effects become less pronounced. For larger system sizes, the transition becomes more sharp. Next, for a fixed system size ($L=128$), the polar order parameter is plotted for different area fractions, ranges from $\phi=0.12$ to $0.48$ and is shown in Fig.\ref{fig:phase_D1}(h). The results indicate that variations in particle density have a negligible influence on the characteristics of the phase transition.

\subsection{Hexatic order}
In this section, we study the extent of hexatic ordering within the system. To this end we define a local hexatic order parameter $\psi_i$ for each particle and a global hexatic order parameter $\psi_6$ as \cite{chaikin1995principles}:
\begin{align}
    \psi_6= \frac{1}{N} \sum_{i}^{N} \psi_{i},
    \qquad
    \psi_{i} =  \frac{1}{{N^n_i}} \sum_{j}^{{N^n_i}} e^{i6\theta_{ij}}.
\end{align}
Here, $\theta_{ij}$ denotes the angle between particles $i$ and particle $j$, and ${N^n_i}$ represents the number of nearest neighbors of particle $i$ (we have used Voronoi tesselation method). The observable $\psi_6$ is obtained by averaging over $1000$ distinct particle configurations in the steady state. The corresponding global hexatic order is presented in Fig. \ref{fig:pis6GR}(a) in the ($n_p,\Lambda_t$) plane. It can be seen from the figure that in absence of pinning,  $n_p=0$, a strong global hexatic order exists. However, the introduction of even a small fraction of pinning sites is sufficient to disrupt this order. The other observables presented in the upcoming sections also support these observations.\\

\subsection{Pair correlation}
We next examine the radial distribution of particles, quantified by the spatial pair correlation function; this is given by: 
\begin{align}
    g(r) = \frac{1}{N} \sum_{i,j} \delta (r - |\mathbf r_{i} - \mathbf r_{j}|).
\end{align}It is computed for the flocking phase at various pinning fractions: $n_p=0$, $n_p=0.01$ and $n_p=0.15$ as shown in Fig.\ref{fig:pis6GR}(b). The reference parameters used in these simulations are $\Lambda_t=0.06$ and $\Lambda_r=0.10$. In the absence of pinning ($n_p=0$), the system exhibits a well-ordered crystalline structure (finite-sized), characterized by pronounced peaks in the correlation function. Introducing a small pinning ($n_p=0.01$) disrupts this ordering, resulting in a transition to a liquid-like phase with diminished peak intensity. Further increasing the pinning fraction to $n_p=0.15$ causes the peak of $g(r)$ to shift toward smaller $r$, indicating a reduction in the minimum interparticle distance.\\

Another interesting parameter to study is the polar pair correlation function $g(r, \varphi)$ \cite{zhang2021active, subramaniam2025minimal}, which explicitly incorporates the orientational dependence of the particles. This is a related quantity to the radial distribution function $g(r)$ to differentiate between spatial structures of different flocks. The angle $\varphi$ defines the relative orientation between two colloids, and is defined as $\cos (\varphi_{ij}) = \mathbf{e}_{i} \cdot \hat{\mathbf r}_{ij}$, where $\hat{\mathbf r}_{ij}=(\mathbf r_{j} -  {\mathbf r}_{j})/| \mathbf r_{j} -  {\mathbf r}_{j}|$ (where $\varphi_{ij}$ denotes the angle between the orientation of the reference colloid ($i$) and the line connecting the centers of the two colloids in the pair ($i$ and $j$)). In Fig.\ref{fig:pis6GR}(c), and (d), $g(r, \varphi)$ is presented with the pinning fraction $n_p=0$ and $n_p=0.01$ respectively, for a system of total $N=4000$ particles. In the unpinned case ($n_p=0$), a well-defined crystalline structure of finite size (referred to as a crystallite flock) is observed, characterized by sharp and intense positional correlations. In contrast, the introduction of a small pinning fraction ($n_p=0.01$) leads to a transition toward a disordered, liquid-like phase, where the correlations are spatially more diffuse. For the crystallite flock, the intensity—corresponding to the probability of finding a particle at well-localized hexagonal lattice sites—is significantly higher compared to the liquid flock, which exhibits reduced intensity and broader spatial distribution. These observations are consistent with the radial distribution $g(r)$ (shown in Fig.\ref{fig:pis6GR}(b)) reinforcing that even a small degree of pinning can induce a pronounced transition from a spatially ordered crystallite state to a liquid-like phase.\\

\begin{figure*}[t!]
    \centering
    \includegraphics[width=\textwidth]{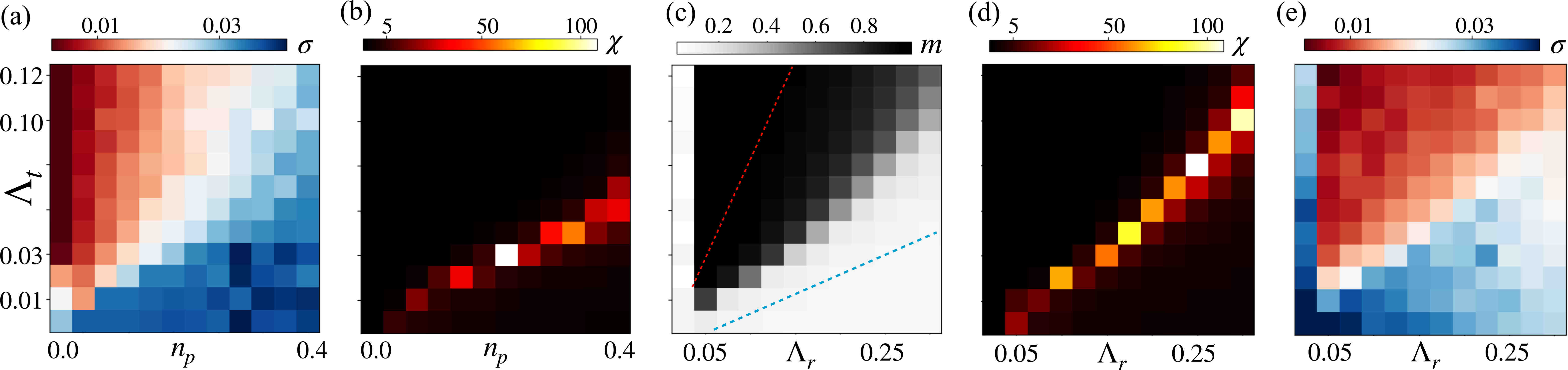}
    \caption{\label{gr}
Phase diagram in ($n_p,\Lambda_t$) plane of spatial density variance $\sigma$ in (a) and susceptibility $\chi$ of polar order in panel (b).
   Phase diagram in the ($\Lambda_r$, $\Lambda_t$) plane: (c) Polar order parameter $m$ with constant $n_p=0.15$. Dotted curves indicate transition line of polar order with $n_p=0$ (marked as blue), and $n_p=0.5$ (marked as red) respectively. (d) Susceptibility $\chi$ of the order parameter with $n_p=0.15$, showing maximum values near the transition line. (e) Phase diagram of spatial density variance $\sigma$ with constant $n_p=0.15$.
    }
\label{fig:fig4}
\label{fig:phase_D2}
\end{figure*}

\subsection{Fluctuations}
Next, we explore variation of different fluctuations in space and orientations. To quantify spatial fluctuations, we first calculate the spatial density variance \cite{knevzevic2022collective}, $\sigma$, derived from the density distribution corresponding to a representative point in the phase diagram, shown in Fig.\ref{fig:fig4}(a). Local density measurements are obtained using a Voronoi tessellation performed at each time step for a given particle configuration. The local density is defined as 
\begin{align}
    \rho_l=\frac{\pi b^2}{A_l}.
\end{align}

Here, $A_l$ denotes the Voronoi cell area associated with each particle. The density variance $\sigma$ is then computed in terms of local density $(\rho_l)$ as: 
\begin{align}
    \sigma  = \langle \rho_l^{2} \rangle - \langle \rho_l \rangle^{2}
\end{align}
It is averaged spatially and subsequently over $1000$ independent configurations. The phase diagram of the parameter $\sigma$ (density variance) is shown in Fig.\ref{fig:fig4}(a) in the ($n_p,\Lambda_t$) plane. In the crystallite phase with $n_p=0$, $\sigma$ is negligible. Within the ordered phase, $\sigma$ remains small but increases near the transition boundary. At higher $n_p$, corresponding to the random phase, $\sigma$ becomes significantly larger.
Next, to characterize orientational fluctuations, we evaluate the susceptibility associated with the polar order parameter \cite{baglietto2008finite,adhikary2021effect}, can then be estimated from the fluctuation in order parameter as
\begin{equation}
  \label{c2d5}
  \chi = L^2 [\langle M^2\rangle - \langle M\rangle^2]
\end{equation}  
where $M$ represents the instantaneous global polar order parameter. Near the flocking transition, the susceptibility $\chi$ exhibits pronounced peaks, indicating enhanced fluctuations relative to both the ordered and disordered regimes. The maximum of $\chi$ thus provides a reliable estimate of the transition point. The computed susceptibility $\chi$ of the system is shown in Fig.\ref{fig:fig4}(b), where the flocking transition line is clearly identifiable (see also Fig.\ref{fig:phase_D1}(a) plot of polar order in the same phase plane). \\

\subsection{ Phase diagram in ($\Lambda_r$, $\Lambda_t$) plane with pinning}
Now, we will study the polarization in terms of translational and rotational repulsion. The pinning fraction is kept fixed to $n_p=0.15$ in this case. The phase diagrams representing polar order, susceptibility and density variance are determined in the ($\Lambda_r,\Lambda_t$) plane and shown in Fig.\ref{fig:phase_D2}(c)-(e). The phase diagram of polarization is shown in Fig.\ref{fig:phase_D2}(c). It is observed that to sustain flocking, $\Lambda_t$ need to be sufficiently high for higher $\Lambda_r$. In Fig.\ref{fig:phase_D2}(d), the susceptibility phase diagram shows the highest intensity near the transition line. In Fig.\ref{fig:phase_D2}(c), the flocking transition line for $n_p=0$ (blue) and $n_p=0.5$ (red), are also shown for comparison. These lines are approximated from the line of maximum $\chi$ values respectively (similar to Fig.\ref{fig:phase_D2}(d)). When there is no pinning disorder present in the system, the flocking phase sustains with a relatively lower value of the parameters $\Lambda_t$ and $\Lambda_r$. Whereas, if half of the population is pinned with $n_p=0.5$, much higher value of this parameter is needed to obtain the flocking phase. A density variance plot with $n_p=0.15$ for the respective phase plane ($\Lambda_r,\Lambda_t$)is shown in Fig.\ref{fig:phase_D2}(e). It is observed that the value of $\sigma$ is low in the polar ordered phase. It starts decreasing around the transition and relatively high in the random phase, particularly with low $\Lambda_r$ and $\Lambda_t$ values.\\

In this section, we have analyzed the results of our model for chemorepulsive active particles subjected to pinning disorder. The particles interact through short-ranged steric repulsion, which prevents overlap, and through long-ranged phoretic forces and torques. As previously reported \cite{subramaniam2025minimal}, in the absence of pinning disorder, the system exhibits a crystallite phase with polar order, referred to as the chemorepulsive crystallite flock (CCF). Our analysis demonstrates that even a small fraction of pinning disorder destabilizes the crystallite flock, leading to the emergence of a liquid flock structure. We now turn our attention to the case where $\Lambda_t=0$, i.e, chemo-repulsive force is absent. In this regime, it was observed that, without disorder, liquid flocks form up to a small but finite value of $\Lambda_r$. Moreover, near the transition to the polar-ordered phase, the system exhibits the formation of density bands. In the following section, we investigate the influence of pinning disorder on this case without chemo-repulsive force present in the system.\\

\begin{figure*}[t]
    \centering
    \includegraphics[width=\textwidth]{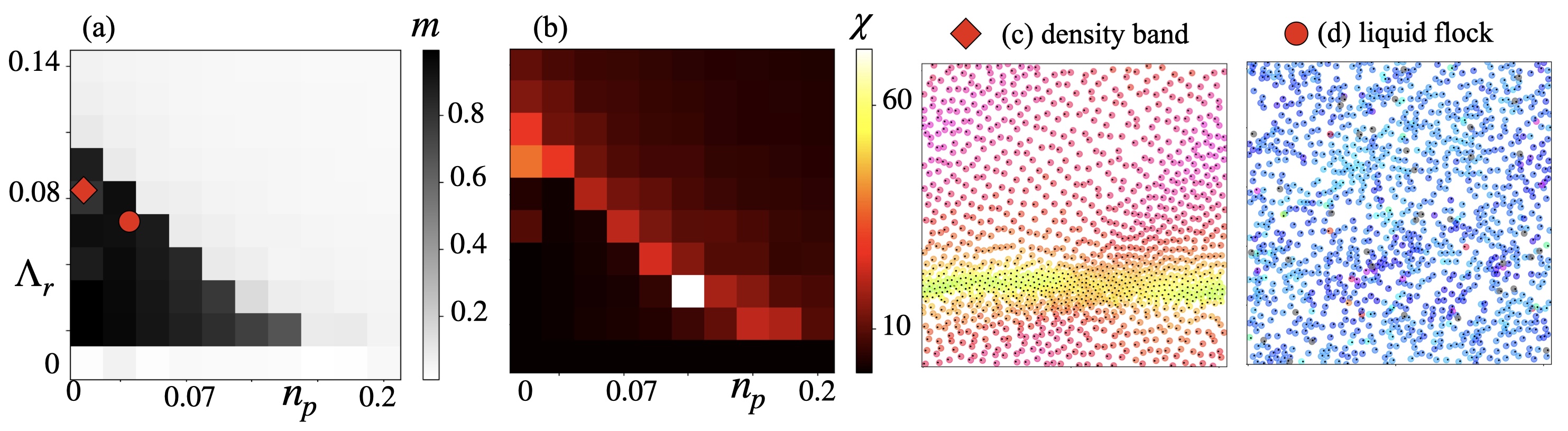}
    \caption{\label{phase_D3}
Case with $\Lambda_t =0$: Phase diagram in the $(n_p,\Lambda_r)$ plane for (a) polar order parameter $m$, (b) susceptibility $\chi$ of the order parameter. (c) density band with $n_p=0$, (d) liquid flock (no band forms) with $n_p=0.03$. The value of the parameters marked in (a) for these two steady-state configurations. The same colour scheme is followed as shown in Fig.\ref{fig:phase_D1}(f). Pinned particles in (d) are marked grey. Total number of particles taken as $N=1200$.  }
\label{fig:phase_D3}
\end{figure*}

\subsection{Effect of pinning without chemo-repulsive force}\label{sec:v}
In this section, we investigate the effect of disorder in the absence of long-ranged phoretic force ($\Lambda_t=0$). Without pinning, this thus constitutes the essential "minimal flock" reported in \cite{subramaniam2025minimal}. We analyze the behavior of the polar order $m$ for various values of the repulsive torque strength $\Lambda_r$ and the pinning fraction $n_p$. The resulting phase diagram in the ($n_p,\Lambda_r$) plane is shown in Fig. \ref{fig:phase_D3}(a). For low pinning fractions ($n_p \approx 0$), the system maintains polar order up to a finite value of $\Lambda_r$ ($\approx 0.1$). However, as $n_p$ increases, the maximum value of the range of $\Lambda_r$ over which polar order persists decreases significantly. Beyond a threshold $n_p \approx 0.16$, the system loses global order, and the collective motion becomes random ($m \approx 0$). The corresponding phase diagram of the susceptibility of the order parameter is presented in Fig.\ref{fig:phase_D3}(b). It is observed that near the transition line, $\chi$ is maximum and exhibits a broader spread toward the disordered phase (right), in contrast to the lower values observed in the ordered phase (left). The system configuration for the unpinned case($n_p=0$) is shown in Fig.\ref{fig:phase_D3}(c). Here, a density band emerges in the ordered phase, near the transition region. Particles within the traveling band move predominantly perpendicular to the longitudinal axis of the density band. These traveling density bands tend to appear in various systems with short-ranged alignment interactions between the self-propelled particles \cite{chate2008collective,tang2025reentrant}. In the present model, when a small fraction of the pinned particles is introduced ($n_p=0.03$), as illustrated in Fig.\ref{fig:phase_D3}(d), the formation of density bands is suppressed due to the influence of pinning disorder and instead a liquid flock structure appears. This effect depends on the finite size of the system. For a larger system with $4000$ particles (as shown in Movie II in the supplementary material), bands still form with a smaller disorder with $n_p=0.01$. However, increasing disorder generally leads to a reduction in the band density, which is also sensitive to system size. Furthermore, in the presence of a chemo-repulsive force and finite disorder, it is possible to observe the formation of density bands in very large systems. An extensive analysis on this suggests an exciting direction for future work.

\section{Discussion and Conclusion}\label{sec:conc}
Collective motion and non-equilibrium phases in active matter systems are strongly influenced by the interplay between interparticle interactions and environmental disorder. In this work, we have investigated the impact of quenched disorder, introduced in the form of pinning sites, on the stability of polar order in a system of chemically interacting active colloids. In our model, active particles interact via long ranged repulsive forces and torques. In a system where pinning is absent, a crystallite flock phase appears at a suitable range of competition between deterministic forces and torques \cite{das2024flocking, subramaniam2025minimal}. A small pinning fraction nullifies this crystallite phase, in place of a polar liquid phase. Relatively high pinning fraction can sustain the flocking phase as long as the repulsive forces are substantial. To further probe the mechanisms underlying these transitions, we examine cases in which specific interaction components are selectively removed. 
% When the short-range volume exclusion potential is omitted, flocking behavior disappears at relatively small values of the long-ranged alignment and repulsive interaction strengths. Conversely, 
When the long-ranged repulsive force is absent, the system retains polar order up to a smaller but finite fraction of pinning sites. \\

In a previous work \cite{subramaniam2025minimal}, minimal mechanisms to obtain crystallite and liquid flocks with repulsive forces and torques were obtained using the current model in the absence of disorder.  The former was referred to as CCF (chemorepulsive crystallite flocks), while the latter was referred to as CLF (chemorepulsive liquid flocks). The pair collision mechanism for the latter was understood in terms of a requisite range of sliding length (along the comoving direction) whilst rotating away due to repulsion; in the former this sliding is further shortened thus giving a crystallite macroscopic structure. We briefly mention here how this pair-sliding mechanism is consistent with the results we report. Firstly, we have shown that CLF formation proceeds the addition of pinning disorder to the CCF; pinned particles can be understood to not contribute to the enhanced repulsion in the pair picture (thus less localization, which is required for the crystal). Note that although pinned particles contribute to the bulk chemical gradients, the enhanced repulsion at the pair collision is suppressed (since they are fixed and hence do not participate in pair collisions). We have separately reported in addition that CLF flocks in the absence of long-ranged translational forces are highly non-robust to pinning; this corresponds to the case of sliding time being too large due to suppressed $\Lambda_r$ on account of the existence of pinned sites (recall that this scales $\propto \frac{b}{\Lambda_r}$ \cite{subramaniam2025minimal}). We note that robustness as discussed here refers to the behavior of the global polar order parameter, and not necessarily other measures; for instance, even a tiny amount of pinning disorder destroys the (hexatic) spatial order of the CCF. \\

Overall, these thus suggest a novel flocking transition induced via a pinning disorder, contrasting from (for instance) findings of network-like flocking topologies seen when there is explicit obstacle-particle interaction \cite{morin2017distortion}. It would be interesting to see if methods of introducing disorder either via synthetic (e.g. microfluidic) obstacles \cite{morin2017distortion, prakash2017trapping}, or via optical traps \cite{gokhale2014growing} may reproduce such a transition; indeed our pinned-active interaction may have to be further tuned in these cases. Potentially important future work would include a detailed study of finite-size scaling of quantities, such as susceptibility, which can tell us about the order of the phase transition and situate our model within a universality class \cite{mutnejaprl2025}. On the applied front, this hints to the existence of an active analogue to so-called ``phase-change materials" \cite{anbarasu2011understanding}, where the dynamic phase is instead tunable.\\

\section*{Conflicts of interest}
There are no conflicts to declare.
\section*{Data availability}
The datasets are generated from computer simulation. They are available from the corresponding authors on reasonable request.
 
\section*{Acknowledgments} 
We thank Professors Hugues Chaté and Chandan Dasgupta for comments. We also thank the two anonymous reviewers for comments, which led to improvement of our manuscript. 
SA acknowledges support from the National Postdoctoral Fellowship (SERB File number: PDF/2023/002096) provided by ANRF, Government of India.

%%%=========================================
%%%=========================================
%%%=========================================
\appendix

%%%=========================================
%%%=========================================

 \section{Table of parameters}\label{app:sim}

A table of all the parameters used for the generated figures is presented in Table \ref{si_table_params}. \\

\begin{table*}[t]
	\centering
	\begin{tabular}[c]{|l|l|l|l|l|l|l|l|l|}
     	\hline
     	Figure no. & $\kappa$ & $N$ & $L$ & $\phi$ & $n_p$ & $\Lambda_t$ & $\Lambda_r$ & $D_r$ \\
     	\hline
     	2.(a)-(e) & $175$ & $1200 $ & $102$ &  $0.36$ & $(0,0.4)$  & $(0,0.12)$ & $0.1$ & $ 10^{-4}$  \\
        \hline
     	2.(g) & $175$ & $(117,7514)$ & $(32,256)$ & $0.36$ & $(0,0.67)$ & $0.05$ & $0.1$ & $ 10^{-4}$  \\
        \hline
        2.(h) & $175$ & $(626,2503)$ & $128$ & $(0.12,0.48)$ & $(0,0.67)$ & $0.05$ & $0.1$ & $10^{-4}$ \\
        \hline 
        3.(a) & $175$ & $1200 $ & $102$ &  $0.36$ & $(0,0.4)$  & $(0,0.12)$ & $0.1$ & $ 10^{-4}$  \\
        \hline 
        3.(b)-(d) & $175$ & $4000$ & $186$ & $0.36$ & $0,0.01$ & $0.06$ & $0.1$ & $10^{-4}$  \\ 
        \hline         
        4.(a)-(b) & $175$ & $1200 $ & $102$ &  $0.36$ & $(0,0.4)$  & $(0,0.12)$ & $0.1$ & $ 10^{-4}$  \\
        \hline
        4.(c)-(e) & $175$ & $1200$ & $102$ & $0.36$ & $0.15$ &$(0,0.12)$ & $(0,0.3)$  & $ 10^{-4}$ \\
        \hline    
        5 & $175$ & $1200$ & $102$ & $0.36$ & $(0,0.2)$ & $0$ & $(0,0.14)$ & $10^{-4}$  \\
        \hline        
	\end{tabular}
 \caption{Parameter values used for respective figures of the paper. 
 Here $L$ is the system size, $\phi$ area fraction related to density, $N$ is the total number of particles and $n_p$ is the pinning fraction. The time step size $dt=0.01$ are kept fixed for all simulations. The remaining parameters are defined after Eq.\eqref{eq:mainLE}. We have kept $\frac{\lambda_0}{4\pi D_c}=1$ in all the simulations. Note that we consider noise $\bm\eta^r$ to be less dominant to deterministic effects. Indeed, for the results in this paper, the Péclet number $\mathrm{Pe} = v_s/(bD_r)$ is around $10^5$. The speed $v_s$ (intrinsic) of a free particle is kept fixed at $v_s = 50$. We have taken the radius of a particle as $b=1$. The total number of particles varies in the range of $N=[117, 7514]$. All the results are reported for $\lambda=0$. We note our results are indistinguishable from the case of a small value of $\lambda$ ($<0.1$). 
 }
 \label{si_table_params}
\end{table*}

\section{Description of the supplementary movies}
The time evolution of the flocking pattern in the presence of a small pinning strength is attached as supplementary movies \cite{siText}. For both cases, we started with random initial conditions (positions and orientations), and update the equation of motion following Eq. \ref{eq:mainLE}. The free particles are colored  by their orientations (given in terms of the angle $\theta$ their orientation vector makes with the positive $x$-axis). The color bar is same as the one used in Fig. \ref{fig:phase_D1}(f). The pinned particles are marked as gray, irrespective of their orientations. We note that our results are robust if we change the system sizes in the simulations.

\begin{itemize}
    \item \textbf{Movie I}.
In the case of $\Lambda_t > 0$, the crystallite structure hindered by few pinned particles. Total number of particles $N=1000$, number of pinned particles $N_p=5$ ($n_p=0.01$).
\item \textbf{Movie II}.
In the case of $\Lambda_t = 0$, density band pattern becomes less prominent with a small number of pinned particles. Total number of particles $N=4000$, number of pinned particles $N_p=40$ ($n_p=0.01$).
\end{itemize}
%%%==================
%%%==================
%%%==================
%%%==================
%%%==================
%%%==================
%%%==================
%%%==================
%%%==================%%%==================
%%%==================%%%==================
%%%==================%%%==================
% \bibliography{reference}
\end{document}